\documentstyle[twocolumn,aps,prl,overcite]{revtex}
\input epsf

\tighten
\begin{document}
%\draft

%\twocolumn[\hsize\textwidth\columnwidth\hsize\csname @twocolumnfalse\endcsname

%%%%%%%%%%%%%%%%%%%%%%%%%%%%%%%%%%%%%%%%%%%%%%%%%%%%%%%%%%%%%%%%%%%%%%%%%%

\title{ 
General considerations of the
cosmological constant and the stabilization of
moduli in the brane-world picture
}

\author{Paul J. Steinhardt
}

\address{Department of Physics \\
Princeton University \\
Princeton, NJ  08544}

\maketitle
%%%%%%%%%%%%%%%%%%%%%%%%%%%%%%%%%%%%%%%%%%%%%%%%%%%%%%%%%%%%%%%%%%%%%%%%%%
\begin{abstract}

We argue that the brane-world picture
with matter-fields confined to
4-d domain walls  and with gravitational interactions across the bulk
disallows adding an arbitrary constant to
the low-energy, 4-d effective theory --
which finesses
the usual cosmological constant problem.
The analysis also  points to difficulties in
 stabilizing  moduli fields; as an alternative,
 we suggest scenarios in which the moduli motion is
 heavily damped by various cosmological mechanisms and
 varying ultra-slowly with time.

\end{abstract}

%%%%%%%%%%%%%%%%%%%%%%%%%%%%%%%%%%%%%%%%%%%%%%%%%%%%%%%%%%%%%%%%%%%%%%%%%%

One of the great mysteries of fundamental physics is to explain the 
value of the cosmological constant, $\Lambda$.   
For a homogeneous and isotropic universe, $\Lambda$
appears as an arbitrary additive 
constant in the Friedmann equation:
\begin{equation}
H^2 =\frac{8 \pi G}{3} \rho  + \frac{1}{3}\Lambda - \frac{k}{a^2}
\end{equation}
where $H \equiv \dot{a}/a$ is the Hubble parameter, $a(t)$ is the 
Robertson-Walker scale factor, 
$G$ is Newton's constant, $\rho$ is the energy 
density, and $k$ is the curvature constant.  In contemporary discussions,
$\Lambda$ has  been interpreted as a vacuum energy
$\rho_v \equiv \Lambda/8 \pi G$.  Dimensional arguments suggest that 
$\rho_v = {\cal O}(M_p^4)= {\cal O}(10^{80})$~GeV, whereas current observational bounds require 
$\rho_v < 10^{-47}$~GeV. To explain the 100 orders of magnitude 
discrepancy, the common point-of-view in fundamental
physics has been that $\Lambda$ is precisely zero
due to some symmetry to be determined.
With recent evidence that the universe may be accelerating,\cite{science} 
the possibility of a 
tiny but non-zero $\Lambda$  is being considered, but there 
is no compelling argument why it should be so.
An alternative possibility is $\Lambda=0$ and the acceleration is due to 
quintessence, a dynamical, time-varying,
spatially-inhomogeneous component.\cite{CDS}

A parallel development has been the  generalization
of string theory  based on branes 
 and the notion of finite  
 ``extra 
 dimensions."\cite{horava,ovrut,dienes,dimopoulos,randall,randall1,randall2}
 For the purposes of this paper, it is not important if the extra
 dimension is large or small,  so long as it is finite.    
The brane-world view suggests that our four-dimensional universe is actually
a slab bounded by two four-dimensional 3-branes separated by a finite
interval in a five-dimensional space.  A particular example is the five-dimensional effective action for $M$-theory on $S^1/Z_2$ obtained by reducing
the 11-dimensional theory on a Calabi-Yau three-fold.\cite{horava,ovrut,luk2,ell1,ell2,luk3}  The five-dimensional
effective theory has the structure ${\cal M}_5=S^1/Z_2 \, \times \, {\cal M}_4$,
with two bounding 3-branes, ${\cal M}_4^{(1)}$ and ${\cal M}_4^{(2)}$.
(The Calabi-Yau
compactified dimensions, presumably small compared to the fifth dimension,
  will not play any important role
in this paper.)
Ordinary matter is confined to this surface, separated by finite 
intervals from one or more 
other surfaces where other particle/fields may  exist.
In one formulation, the only interaction in the bulk is gravitational
in nature.  Ordinary matter  is described by fields confined to one
of the branes, hidden matter is described by fields confined to 
the other brane, and the two sets of fields only interact by 
gravitational interactions that are exchanged across the intervening
bulk.  In other formulations, interactions are universal so other 
particles and fields can be transmitted across the brane.
If the additional fields are very massive (or order the Planck mass) and give
big masses to the moduli fields (thereby pinning the interbrane 
separation at some specific value), then the setup is
 indistinguishable from a general 4-d theory at low energies and
the considerations in this paper may not apply.  
Otherwise, the additional fields make only minor modifications to our
argument.  For simplicity,
we will assume the first formulation, nonuniversal interaction with gravity
only between the branes.

The  expectation is that, at low energies and large distance 
scales compared to the compactified dimensions, the brane-world 
picture should reduce to a  4-d, low-energy, effective
 field theory.
 In most respects, the theory is equivalent to a 
true 4-d theory with ordinary and hidden matter fields existing on the same
$3+1$-dimensional space.
The purpose of this paper, though, is to point out that  genuinely 
new effects
occur by separating fields onto two planes separated by a fifth dimension.
The effects depend directly only on the 
properties of the brane-world setup without explicit dependence on
supersymmetry. They have an impact on our understanding of the cosmological
constant and may enable a cosmological solution to 
the problem of fixing the moduli fields.

We begin by  posing a simple argument 
that {\it the 
brane-world view of the universe is incompatible 
with a ``true cosmological constant" in the projected, 
effective 4-d field theory; unlike the case of an ordinary 4-d theory,
 one does not have the freedom to add an 
arbitrary  $\Lambda$
to the Friedmann equation (Eq. (1)) or an arbitrary constant to the action 
without also introducing  interactions with other fields in the 4-d theory.} 
As a result,  the 
expansion of the universe is dependent on the expectation values
of those fields and not just the constant itself.
 When all interactions are taken into account,
this may behave like quintessence and lead to negative pressure and cosmic 
acceleration, but the situation is qualitatively 
different from adding an arbitrary 
vacuum energy only.

The key point is that,
if  vacuum density is  added to one brane,  
this creates a repulsive gravitational
field that acts
in all five dimensions.  In particular, the equivalence principle
in 5-d demands that the gravitational field generated by the vacuum energy on
one brane act on the matter-energy on the other brane, either attracting or repelling it.  Projected into the 4D effective action, the 5-d gravitational
interaction with the other wall becomes a 4-d interaction depending on the 
moduli fields (whose expectation value
determines the distance between branes in 5-d).  Hence, embedding a 4-d theory in five dimensions means that there is no way to add an arbitrary constant $\Lambda$ alone to the projected 4-d theory.  The same conclusion would be reached if the vacuum energy resided on the other wall or in the bulk.

Planar 2-d domain walls in
$3+1$-dimensions 
with positive energy density (and negative pressure)  induce repulsive, gravitational
interactions.\cite{vilenkin,ipser}  If the energy momentum tensor is  $$T_{\mu}^{\nu}=
{\rm diag}(\sigma, \, -p_1,\, -p_2, \, -p_3) =
\sigma  {\rm diag}(1, \, 0,\,-1, \, -1),$$
where $\sigma$ is the surface energy density, then the metric outside
the walls is
\begin{eqnarray}
ds^2 & = &  (1 -2 \pi G \sigma x)^2 dt^2 -dx^2 - 
\\  & & 
(1- 2 \pi G \sigma x)^2
{\rm e}^{2 \pi G \sigma t}( dy^2 + dz^2).
\end{eqnarray}
The $(x,\, t)$ part of the metric is the (1+1)-dimensional Rindler
metric describing flat space in the frame of reference of a uniformly
accelerated observer. (The $(y,\, z)$ planes have de Sitter metric.)
A similar situation arises if vacuum energy $\rho_v$
resides on a 4-dimensional brane 
in 5-d.\cite{kaloper,huey} 
Supposing that one wall contains predominantly vacuum energy and the other contains matter-fields (massive particles), 
the branes repel one other such  that they accelerate away 
from one another  with acceleration $4 \pi \rho_v/3 M^3$, where $\rho
$ is the energy density and $M$ is the 5-d Planck mass.
As recently shown by Huey,\cite{huey} replacing the vacuum energy with a negative pressure component with $p+(2/3)\rho<0$ also produces repulsion, whereas $p+(2/3)\rho>0$ produces an attractive force on the particles on the other wall.  
If both walls contain vacuum energy, the walls also accelerate away from
one another.
(Negative vacuum walls have positive pressure and  are attractive.)
For
other forms of energy with highly negative pressure, the situation is more complicated.\cite{luk3,huey}
If constant vacuum  energy is added to the 
bulk, this also leads to expansion of
the extra dimensions and effective repulsion between branes.

Projected into the low-energy, 4-d effective theory, the gravitational repulsion
between branes in 5-d must be reinterpreted 
as a direct interaction between $\rho_v$, the modulus field $\phi$  and the 
hidden sector fields in 4-d.  
To make the claim precise, it is useful to choose a specific metric 
form for the action so that it is unambiguous what is meant by 
a constant vacuum energy, $\rho_v$.
For this purpose, the projected theory, which may 
include non-minimally coupled fields similar to Brans-Dicke theory,
 ${\cal L} = f(\phi) R + \ldots$,  
should be Weyl-transformed into standard Einstein form,
${\cal L'}= R +\ldots$. 
For an ordinary,    4-d quantum field   theory, there 
is no symmetry to forbid adding to ${\cal L'}$
an arbitrary constant, $\rho_v$ (and 
introduce no other terms).\cite{brans}  This
term would only add  a constant  in 
the Einstein-frame 
Friedmann equation, $\Lambda = 8 \pi G \rho_v$, and trigger exponential 
expansion; the freedom to add this constant is what underlies the 
classic cosmological constant problem.
  In
the brane picture, 
though, adding the constant term alone -- a ``true" cosmological
constant -- is not possible since vacuum energy introduces effects
in the extra-dimensions which project into additional interactions
in ${\cal L'}$.\cite{foot2}

The nature of the interactions in the 4-d 
effective theory induced by $\rho_v$ 
can be guessed from the brane-world picture.
In ordinary field theories,
the value of $\Lambda$ is $ 8 \pi G_4 \rho_v$ where 
$G_4$ is the 4-d Newton's constant.  In the brane-world picture,
$G_4$ is inversely proportional to the distance between 
branes which, in turn, is proportional 
to the 
orbifold modulus  $T = {\rm exp}(\phi)$. In the 4-d theory,
$\phi$ reduces to a quantum field and $\Lambda$  is proportional
to a monotonically decreasing function of $\phi$,
rather than a constant.
Because of the repulsive interaction between branes, there is
also a $\phi$-dependent interaction between the vacuum energy
on one brane and  matter-energy fields on the other brane.\cite{foot}
%%%
Another  interaction is  due to
fluctuations in the surface of one brane  which create
fluctuations in the distance between the two branes as one scans
across the 3-brane and, hence, 
(4-d) spatial dependence of the interaction
resulting couplings between $\rho_v$ and gradients in $T$.

A related effect occurs if the brane-world view is placed in 
a more general  cosmological setting.  A matter or thermal distribution of 
energy on our 4-d brane induces an attractive gravitational
force between
branes\cite{huey}  or, equivalently, an interaction  between the matter or
radiation density and the modulus field.
The interbrane gravitational
force in 5-d, proportional to $\rho+(2/3) p$,\cite{huey}
must reduce in the low-energy,
4-d theory to an interaction  with  factors proportional to 
$\rho_i+(2/3) p_i$, where $\rho_i$ and $p_i$ are, respectively, the energy density and pressure on a each brane, $i$.
(N.B., if the universe has recently transformed from 
being matter-dominated to being  $\Lambda$-dominated, as suggested by some recent observations,  the force
has switched from attractive to repulsive. This should induce
at least some small change in the modulus and, therefore, the 
gravitational constant.)

It is interesting to  compare our conclusion
to the behavior of 4-d supergravity (SUGRA) field theories.
Adding 
a constant alone  to the SUGRA potential energy density is disallowed because it
 explicitly breaks supersymmetry.  Instead, a constant can
be added to the superpotential from which the potential is 
computed. The constant term in the superpotential 
induces additional interactions with all fields in the potential. 
We have drawn 
a qualitatively similar conclusion based on  the brane construction 
even though no explicit dependence on supersymmetry is required.
 
The gravitational interaction between branes produced 
by matter, radiation and vacuum energy suggests some insights
into the stabilization of moduli and supersymmetry breaking.
The stabilization of the $T$ modulus field corresponds to fixing
the distance between branes in some static ground state where
 $T$ is set
at the minimum of its effective potential.
However, if the force between branes is purely gravitational, this
option does not seem possible. 
Depending on the sign of the  vacuum density on
the branes, the gravitational force between branes
is either attractive,  repulsive, or neutral,
but always  with the same dependence on the modulus field. 
 The  
problem of fixing $T$, therefore,  becomes acute since it does
not seem possible to combine monotonic forces of the same
type so as to produce an effective potential with a  stable minimum.
Examples of static brane  solutions can
be constructed\cite{ovrut,randall}  by having a mix of vacuum 
densities with different signs on the branes and bulk;
the solution is achieved by carefully canceling 
attractive and repulsive forces.
However,
the configuration is unstable since 
a slight perturbation in $T$ will cause the repulsive or attractive
force to dominate and the walls to fly apart
or fall towards one another.  (The moduli stabilization problem is, in this sense, reminiscent of Einstein's problem of making a static universe. The tendency of energy to attract or repel gravitationally means that a static solution can only be found by carefully balancing the two and, then, even if accomplished, 
the solution is unstable.)
When the model is made realistic by adding matter and radiation
on the walls, then stabilization of the modulus is further confounded by the
added  (attractive) gravitational force between walls due to 
these time-varying energy densities which drive the universe away from the static solution.

Only a few options for  stabilization seem to be available. 
One possibility is that
there are additional charges on the walls and 
additional  fields besides the graviton
being exchanged
 across the 
bulk which produce  a non-gravitational force with different distance
dependence. (As discussed in the introduction, we only consider
the case where the additional, exchanged fields are light compared to the
compactification scale.)
For supersymmetric BPS $d$-branes, exchanges of scalar and 
3-form fields contribute canceling interactions such that the the net
force between branes is zero, independent of 
distance.\cite{ovrut} 
This particular example  does not seem
very satisfactory since it relies on exactly supersymmetric ground
states, and the net force is zero rather than stabilizing.
However, perhaps some solution of this type can be found in the 
supersymmetry breaking case.
Another recent suggestion\cite{randall2} is that the modulus need not
be stabilized at any finite value. Consistent theories might
be constructed with a semi-infinite distance between walls
by allowing  a metric with non-trivial dependence on the fifth 
dimension.

We wish to suggest an alternative approach in 
which the moduli fields are not absolutely
stabilized but, instead,
change exponentially slowly with time. 
This approach relies on 
placing the model in a realistic cosmological
setting where the finite matter-energy densities on the walls can
play various dynamic roles in damping modular field motion.
In the brane
picture, this means that we accept the possibility of 
imbalanced forces 
between branes, but we also imagine various processes that 
 slow the brane motion.

One candidate for the frictional damping
force is Hubble red shift, an effect which can work equally well for 
$T$ moduli, $S$ moduli and other moduli. Consider the low-energy, 
effective 4-d field theory. A canonical scalar field $\phi$
in 3+1-dimensions
satisfies an equation of motion:
\begin{equation}
\ddot{\phi} + 3 H \dot{\phi}= -V'
\end{equation}
where dot denotes the time derivative,
prime denotes the derivative with respect to $\phi$, 
and $H^2 = 8 \pi G \rho_{tot}/3$ is the Hubble parameter for total energy density $\rho_{tot}$.
If $\phi$ is the modulus field, then all of the 5-d gravitational interactions between energy on the walls and 
the modulus, as described above, are projected into 4-d and incorporated as  field-theoretic interactions in $V$. 
If $V'/\phi \ll H^2$, then the Hubble damping term dominates the potential
term and the field moves exponentially slowly down the potential.  (Readers
will recognize this as the slow-roll phenomenon invoked in inflationary
models.\cite{AS})   A useful figure of merit is $\tau = \dot{\phi}/\phi H$,
which measures the percentage change in the expectation value of 
$\phi$ in one Hubble time.  For the scalar field in the slow-roll
limit, $\tau \approx V'/\phi H$;
hence, if this ratio is sufficiently small, $\phi$ rolls negligibly 
over the lifetime of the universe despite the fact that it lies at
an unstable point in the potential.  
The moduli expectation values may be unstable but Hubble damping can
cause their motion to be exponentially slow with time constant 
 $V'/\phi H$. If variations in $\phi$ induce  variations in
 coupling constants, then this damping can be made small enough to 
 satisfy observational constraints on time-varying
 constants.\cite{Carroll}
 (If the modulus field is too light, there can also be 
 unacceptable fifth force interactions which must be suppressed by
 some separate mechanism.)
 
Hubble damping relies on the gravitational effect of the total energy
density $\rho_{tot}$.  Two modes of damping the field are possible.\cite{HS}
The field may be ``self-damping," a condition which occurs if 
the potential $V$ at a given value of $\phi$
is so flat that 
$V'/\phi V < 8 \pi G/3$. 
An example
is $V \sim {\rm exp}(- \beta  \phi/M_p)$   for  $\phi/M_p \gg  \beta
\sqrt{64 \pi^2}$ ({\it i.e., small $\beta$ or large $\phi$}),
where $M_p^2= G^{-1}$ is the Planck mass.  
In this case, the potential energy for the 
field is sufficient to overdamp the motion  since we have
\begin{equation}
\frac{V'}{\phi} 
< \frac{8 \pi G}{3} V < \frac{8 \pi G}{3} \rho_{tot} \equiv H^2, 
\end{equation}
which is the Hubble damping condition. For this potential,
if the field is Hubble damped 
for a given  value of $\phi$, it continues to be damped forever
as it slowly makes its way down the potential.
 The second mode is where the
field is not self-damping,  ($V'/\phi V > 8 \pi G/3$), but 
$V'/\phi \ll \frac{8 \pi G}{3} \rho_{tot} \equiv H_0^2$, 
where $H_0$ is the present-day value of the Hubble parameter.
A pertinent example is the non-perturbative potential generated
for the dilaton and $T$-modulus field, $V \sim {\rm exp}(-\beta
{\rm e}^{\phi})$, for $\beta > 10$, which is not self-damping for
any $\phi>0$.
In this case, the field motion is Hubble damped by the 
energy density in other components besides $\phi$.  Since $H > H_0$
in the past, the field would have been damped in the past, as well; 
however, it may be that $H$ decreases in the future such that
$V'/\phi V > H^2$, at which point the field would become undamped
and begin to roll at a significant rate.
There remains the challenge of making the field energy small compared to
the present-day matter density and, at the same time, achieving a 
sufficiently large supersymmetry breaking scale.    
The same challenges exist if the field were stabilized but, perhaps
they can be more easily resolved in the damped case. We consider
some specific models in a separate publication.\cite{joint}

A second approach for slowing the moduli field and brane motion is 
through  non-minimal coupling, nonlinear gravity
or non-linear sigma model type 
kinetic energy. Each can recast as 
an added damping term that can
exceed the Hubble damping. For example, 
for a scalar field with action ${\cal L} = 
g(\phi) (\partial_{\mu} \phi)^2$, $\phi$ can be infinitely
damped if $g(\phi)$ has a pole at some $\phi=\phi_0$.
(Here we assume that the canonical kinetic energy has positive 
coefficient and, hence, $g(\phi)>0$.)  The pole, which can arise
from various nonlinear interactions, may only be apparent after
field redefinitions and metric transformation.  For example,
consider,  a scalar field $\phi$ with 
non-minimal coupling to the Ricci scalar (${\cal R}$)
and canonical kinetic energy: the action is ${\cal L} =
- f(\phi) {\cal R} + \frac{1}{2}(\partial_{\mu} \phi)^2 + \ldots$.
We can assume  $f(\phi)$ to be analytic and positive for all $\phi$.
Setting $\Phi = f(\phi)$, the action can be recast in   Brans-Dicke
like form:  ${\cal L} = - \Phi {\cal R} + (\omega(\Phi)/\Phi) 
(\partial_{\mu} \phi)^2 + \ldots$, where $\omega$ is the standard 
Brans-Dicke parameter, $\omega(\Phi) = f(\phi)/2 [f'(\phi)]^2$. 
The limit $\omega \rightarrow \infty$  corresponds
to  Einstein gravity where $\phi$ is constant.
Now we see that, recast in this form, the coefficient of the kinetic
energy becomes singular where $f(\phi)$ has an 
extremum, $f'(\phi)=0$, even though $f(\phi)$ itself has no
singularity.   
(This method of freezing non-minimally coupled scalar fields 
was exploited previously in the context of ``hyperextended 
inflation."\cite{hyperex}) 
A similar effect occurs in nonlinear gravity models
for which the gravitational action is $f({\cal R}) R$, which can
be recast as standard Einstein gravity with 
a non-minimally coupled scalar field.
This approach can freeze the field even when the curvature of the
effective potential (and, hence, the mass) for the modulus is large,
and, thereby, evade both constraints on time-varying constants and 
fifth force.

A third approach takes advantage of the non-linear couplings of the 
moduli fields to the kinetic and potential energies of matter fields
on the boundary.  Although the moduli potential is flat (perturbatively)
in the absence of matter and radiation, a non-zero potential is induced
in a cosmological setting with finite matter and radiation densities.
Depending on whether the matter fields are 
gauge fields, fermions, or scalars, 
the moduli couple in different
ways to the kinetic and potential energy densities (due to different
dependencies on the Kahler potential).\cite{luk4}
That is, if $\psi$ is a matter scalar field, $F^{\alpha}_{\mu \nu}$ is the
field strength 
for  gauge fields on a wall,
and $\phi$ is a modulus field,  the action takes the form
${\cal L} = \ldots + f(\phi) (\partial_{\mu} \psi)^2 - g(\phi)V(\psi)
+ h(\phi) \, tr \,F^2   + \ldots$, where $f$, $g$ and $h$ are different
functions of the modulus field.
In particular, the moduli-dependent coefficients of the kinetic 
energy terms are increasing functions of the modulus for some fields
and decreasing for others. 
See, for example, Eq. 2.3 
in Ref. 14.
As a result, in 
a radiation dominated phase, say,
where there is  a mixture of excitations of these fields,
 a potential for the moduli fields is induced which can
 bound  the moduli. The moduli run off to infinity in the 
 vacuum solution when non-perturbative interactions are included,
 but here the radiation-induced potential dominates the 
 non-perturbative contribution.
   As the universe and expands the energy densities
change, the shape of the potential and perhaps the minimum may change;
if so, perhaps this effect must be combined with damping 
to  freeze the field  enough to meet
observational constraints. We explore this possibility
for a realistic model in a separate paper.\cite{joint}

A final, more speculative approach is to combine two types of 
gravitational interactions between walls to constrain their separation.  
We argued that stabilization is difficult since the gravitational force 
between walls combines attractive and repulsive contributions with the same 
dependence on the interwall separation.  This argument assumed 
homogeneous distributions of stress-energy on either wall. 
But, consider, for example, a case where vacuum energy on either wall 
creates a repulsive force, but the universe also contains 
a finite density of mini-black holes whose Schwarzschild radius 
exceeds the interwall spacing.  In the 5-d theory, the black holes are not 
constrained to lie within one wall because their gravitational field can 
penetrate the bulk and  stretch across to the other wall to form a "black 
string."\cite{flamme}   The black holes produce a local attractive force 
between branes that, in essence, glues the walls together at
certain points. If the dark 
matter, say, consists of these mini-black holes, then the density would be 
rather high within our Hubble volume.
 As the universe expands and the density of black holes 
decreases, the vacuum energy would eventually dominate the universe and the 
walls would recede from one another, ultimately opening up the fifth 
dimension. 
(If the universe is entering a period of accelerated 
expansion in the present 
epoch, as suggested by some recent observations, then this opening of the 
fifth dimension may not be all that distant.)
At present, this approach is only a conceptual picture since we do not 
have a calculational scheme for computing the evolution of the  
inhomogeneous setup.

Stabilizing the modulus and supersymmetry breaking are sometimes thought to
be linked. To construct a supersymmetry breaking state which is
stable, minimal energy, has small or zero cosmological constant, and 
which satisfies the phenomenological constraint on the supersymmetry
breaking scale has proven to be very difficult. Most examples 
seem contrived.
 The  damping concept  suggests that the universe
need not lie at a true ground state nor even at a metastable state.
In this case, the model need not even have a supersymmetry breaking
ground state; if the moduli lie at values away from the ground state,
supersymmetry breaking is generic.
The constraints on model-building are different, and
perhaps simpler to satisfy.  
We will report on some attempts in a separate publication.\cite{joint}
Another application of the brane-world picture is
inflation, in which the hyperextended  mechanisms for starting and stopping 
inflation\cite{hyperex} appear to have a natural reinterpretation in terms
of brane interactions, as well be described elsewhere.\cite{extend}

The author thanks  B. Ovrut and  D. Waldram 
for many conversations and patient tutoring in $M$-theory and 
Tom Banks, R. Caldwell, L. Randall
and   E. Witten for useful discussions.
This research was supported by the US Department of Energy grant
DE-FG02-91ER40671 (Princeton).


\begin{thebibliography}{horava}
\bibitem{brans}  For non-minimally coupled fields, as in Brans-Dicke
theory, the term ``cosmological constant" is sometimes used to refer
to a constant term added to  the action 
where the gravitational action  is in Brans-Dicke (rather than
Einstein) form.  This constant is different from the 
``true cosmological constant"
({\i.e.}, a constant term  that leads to de Sitter expansion)
considered in this paper.
\bibitem{foot}
In 5-d, the force produces a constant acceleration,
$G_5 \rho_v$. where $G_5 = M_5^{-3}$ is the 5-d Newton's 
constant and $M_5$ is the 5-d Planck constant.
The acceleration would derive from an energy density 
proportional the acceleration times the distance between branes,
$R_{orb}$ (the orbifold radius). In the Horava-Witten model, 
the orbifold radius is proportional to $T S^{-1/3}$ (D. Waldram,
private communication), where $T$
and $S$ are the orbifold and Calabi-Yau
moduli fields, respectively.
Since the 4-d Newton's constant
$G_4$ is proportional to $G_5/R_{orb}$, the  interaction between
walls can be reinterpreted as an energy density proportional
to $-G_4 \rho \rho_v R_{orb}^2 \propto T^2 S^{-2/3}$, where $\rho$
is the energy density on the opposite brane.
(The minus sign is due to the fact that the force is repulsive.)
\end{thebibliography}
\end{document}